\documentclass[aps,twocolumn,amsmath,amssymb,showpacs,floatfix,prb]{revtex4}
\usepackage{graphicx}

\newcommand{\etal}{{\it et al.}}
\newcommand{\g}{u}
\newcommand{\beq}{\begin{equation}}
\newcommand{\eeq}{\end{equation}}
\newcommand{\bea}{\begin{eqnarray}}
\newcommand{\eea}{\end{eqnarray}}

\begin{document}

\title{Gap anisotropy and universal pairing scale in
a spin fluctuation model for cuprates}
\author{Ar. Abanov$^1$, A. V. Chubukov$^2$ and M. R. Norman$^3$}
\affiliation{
$^1$ Department of Physics, Texas A\&M University, College Station, TX  77843\\
$^2$Department of Physics, University of Wisconsin, Madison, WI  53706\\
$^3$ Materials Science Division, Argonne National Laboratory, Argonne,
IL 60439}
\date{\today}

\begin{abstract}
We consider the evolution of $d_{x^2-y^2}$ pairing, 
mediated by nearly critical 
spin fluctuations, with the coupling strength.
We show that the onset temperature for pairing, $T^*$, smoothly 
evolves between weak and strong coupling, 
passing through a broad maximum at intermediate coupling.
At strong coupling, $T^*$ is of order the magnetic exchange energy $J$.
We argue that for all couplings,
pairing is confined to the vicinity of the Fermi surface.  
We also find that thermal spin fluctuations
only modestly reduce $T^*$, even at criticality, but they substantially smooth the gap anisotropy. The latter evolves with coupling, being the largest at weak coupling. 
\end{abstract}


\maketitle
{\it Introduction}~~~
Understanding the origin of the 
pseudogap in high-$T_c$ cuprate superconductors is a 
 key problem~\cite{review}.  Some argue that the pseudogap 
  originates from (quasi) long range order in a non-pairing
 channel (two-gap scenario)~\cite{two-gap}. Others
 argue instead that the pseudogap is a
 phase in which fermions already form
 singlet pairs, but long-range superconducting coherence is not yet 
established (one-gap scenario)~\cite{one-gap,mikearcs}. 

The theories within the one-gap scenario can be broadly separated into 
two classes -- `strong coupling' theories which consider a doped
Mott insulator, and `weak coupling' theories which assume a normal metallic state
at large dopings.
It is widely accepted that the cuprates  display a crossover from Mott-like behavior in the underdoped
regime to Fermi liquid-like behavior in the overdoped regime, with the superconducting dome straddling these two regimes.
Quantitatively, strong and weak couplings are the limits of small and large
 values of the dimensionless coupling $\g$, which scales as $U/W$, where $U$ is the effective 
 Hubbard  interaction, and 
 $W$ the bandwidth.  For $\g <1$, it is natural to 
assume that pairing is confined to near the Fermi surface
%AC 
(FS)
% and replaced Fermi surface -> FS below
 and can be thought of as mediated by 
a bosonic `glue', the most natural candidate being collective excitations in 
 the spin channel, enhanced around ${\bf Q} = (\pi/a,\pi/a)$ (i.e., antiferromagnetic 
spin fluctuations).
For $\g >1$, it has been argued~\cite{phil} that the notion of a bosonic glue is 
meaningless since
the dynamics of the superexchange interaction, $J$, occurs only for energy scales of order $U$, but this picture has been  challenged based on simulations of the Hubbard model~\cite{doug}.

\begin{figure}
\includegraphics*[width=0.7\columnwidth] {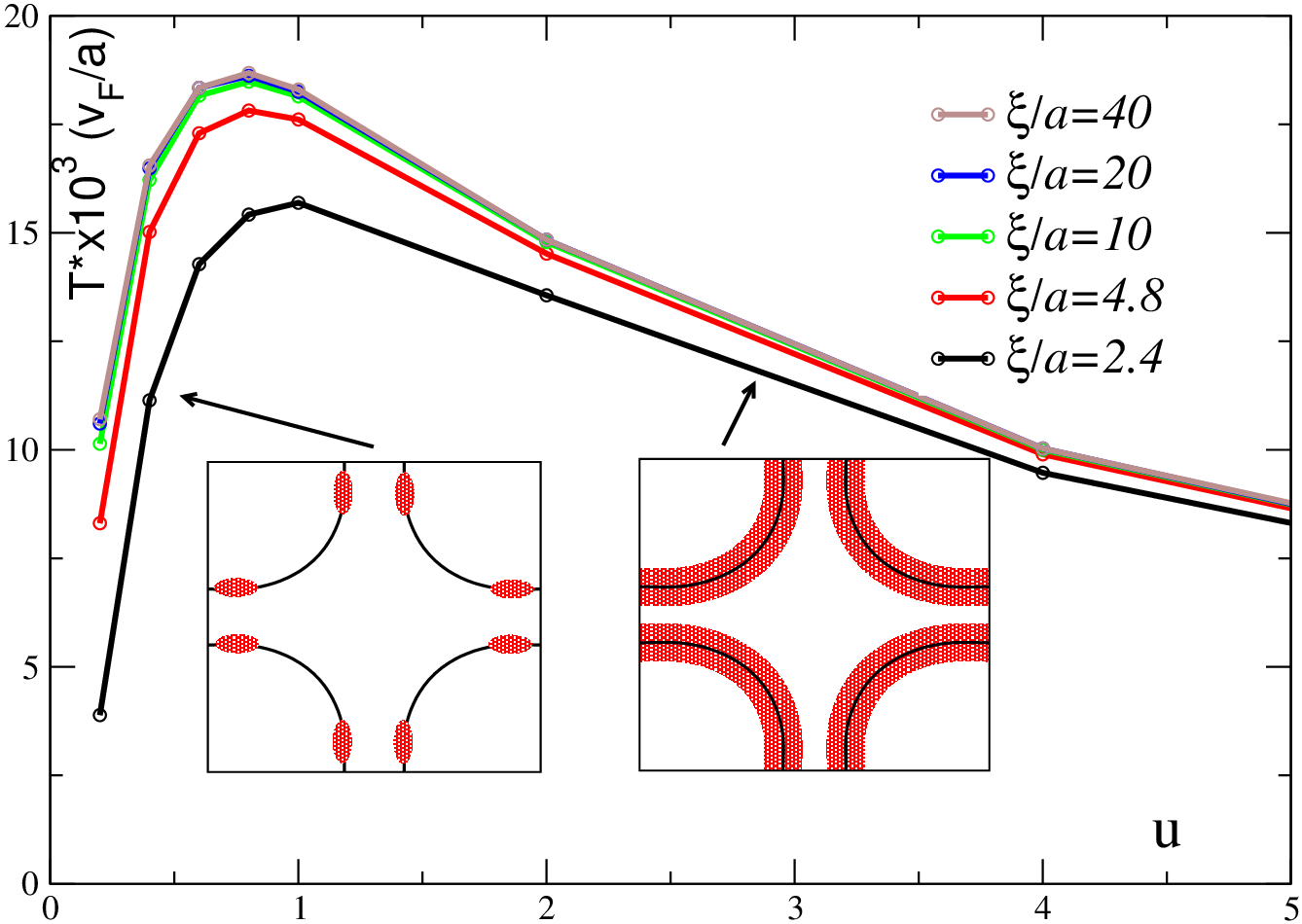}
\caption{(Color online)  The onset temperature  $T^*$ for $d_{x^2-y^2}$ pairing
 vs the dimensionless coupling $\g$ 
for an Ornstein-Zernike form for the static spin propagator near $Q$
 and various values of the magnetic correlation length $\xi$,
with $v_F/a = 1 eV$.
The insets show where the relevant fermions are located along the FS.
For both small and large $\g$, pairing is confined to its vicinity.}
\label{fig1}
\end{figure}

In this paper, we argue that the nature of the pairing is similar
 for both small and large $\g$. We show that the onset temperature for pairing, which is the
 pseudogap $T^*$ in a one-gap scenario, 
 smoothly evolves between  weak and strong couplings, passing through a broad maximum
 at intermediate coupling, where $T^*$ is a
 fraction of the Fermi 
 energy (Fig.~\ref{fig1}). At large $\g$,  $T^* \sim J$. 
 Still, we find that even in this limit, pairing is confined to the vicinity of the FS.
 The only real difference between weak and strong coupling is 
 the range of the FS momenta involved in pairing -- for $\g <1$, pairing comes 
 from regions around the hot spots (Fermi momenta connected by $Q$),
 while for $\g >1$ the whole FS is involved in pairing. We show that $T^*$ weakly 
 depends on the magnetic correlation length $\xi$ and 
can easily reach $300-500K$ for $\g  \sim O(1)$.

We also discuss the special role of static thermal fluctuations,
which scatter with zero energy transfer
and therefore act as non-magnetic impurities that are 
pairbreaking for $d_{x^2-y^2}$ symmetry~\cite{imp,msv}.
Static fluctuations are particularly relevant for
$\xi = \infty$, when their contributions  to the mass renormalization and the pairing vertex diverge. Earlier studies~\cite{msv} suggested that $T^*$ should vanish for $\xi = \infty$. We found that the effect is less drastic than originally thought --
We found that static thermal fluctuations do reduce $T^*$ somewhat, but $T^*$
still remains finite for $\xi = \infty$. 

Finally, we discuss the angular dependence of the $d_{x^2-y^2}$ gap. We find that for small $\g$, 
the gap is very anisotropic and rapidly drops in magnitude upon deviation from the hot spots.
At large $\g$, the gap near the node is actually larger than the $\cos(k_xa)-\cos(k_ya)$ form.
This behavior can most simply be understood from the fact that the anisotropy of the pairing vertex $\Phi$
and the mass renormalization $Z$ approximately cancel in the gap parameter $\Delta \equiv \Phi/Z$
when $Z$ is large.
More formally, static  thermal fluctuations tend to smooth the gap anisotropy in order to
minimize pairbreaking.

{\it The model}~~~~We consider an approach to pairing from the Fermi liquid region of large dopings. We assume that the 
strongest fermion-fermion interaction is  in the spin channel
 for momentum transfers near $Q$. The low-energy physics of such a Fermi liquid is captured by a 
 spin-fermion model which reduces the interaction between low-energy fermions to the exchange
 of two-particle collective modes in the spin channel. 

The spin-fermion model has been described in detail in earlier  
work~\cite{adv}. The inputs are the 
Fermi velocity $v_F$, the spin-fermion coupling $U/2$, and the static boson propagator 
$D_q (\Omega =0)$. In earlier work, $D_q (0)$ was assumed to have an Ornstein-Zernike (O-Z) form 
$D_q^{-1}(0) = \xi^{-2} + |{\bf q}-{\bf Q}|^2$. We will use 
this form of $D_q (0)$, and also a related form $D_q^{-1} (0) =
 \xi^{-2} + |{\bf q}-{\bf Q}|^2 + b ( (q_x-\pi/a)^4 + (q_y -\pi/a)^4)$, with $b>0$,
to model inelastic neutron scattering experiments that show that the spin fluctuations decrease faster
with deviation from $Q$ than the O-Z form predicts~\cite{neutron}.
  
The effective four-fermion vertex $\Gamma(q) = (U/2) D_q$ 
gives rise to fermion and boson self-energies, and to an attractive pairing interaction in the 
$d_{x^2-y^2}$ channel~~\cite{comm_aa}.
In Eliashberg theory~\cite{eliash}, which has been justified earlier~\cite{andy,adv}, the linearized gap equation  has the form 
\beq
\Delta_k (\omega) =
- \frac{\g T}{a} \sum_{\omega'} \oint d k' \nu_{k'} \frac{\Delta_{k'}
(\omega') + \Delta_k (\omega) \frac{\omega'}{\omega}}{|\omega'|}
D_{k-k'} (\omega-\omega')
\label{1}
\eeq
where $\omega$, $\omega'$ are Matsubara frequencies, the integration over $k'$ is along the FS,  $\nu_k$ is the density of states normalized such that  $\nu =1$ along the nodal direction, 
and $\g = 3 U a/(8\pi v_F)$ is the dimensionless coupling where  $v_F$ is the nodal velocity.
The full dynamical interaction $D_q(\Omega)$ 
includes the boson self-energy, which in Eliashberg theory
 reduces to the Landau damping term 
\beq
D_q^{-1} (\Omega) = D^{-1}_q (0) + \frac{|\Omega|}{\Gamma},~~\Gamma = 
\frac{3a}{16} \frac{v_F}{\g}
\label{2} 
\eeq  
We emphasize that the boson self-energy comes from the same interaction that gives rise to 
pairing,  
and  $\Gamma$ contains the same dimensionless coupling $\g$.

%AC restored
The overall   sign in the r.h.s.~of Eq.~(\ref{1}) is 
a consequence of the fact that pairing is in the spin channel.
 Isotropic $s-$wave pairing is not possible in this case,
 while $d_{x^2-y^2}$ pairing is 
favored because $\Delta_{k} (\omega)$ and $\Delta_{k'} (\omega)$ 
have opposite sign for ${\bf k}' = {\bf k} + {\bf Q}$, where $D_0 (q)$ has 
a maximum~\cite{comm_aa}.

%AC
Eq. (\ref{1}) has a solution at $T = T^*$. As the
 temperature only appears explicitly in Eq.~(\ref{1}) in the Landau damping term $\Omega/\Gamma = 2\pi n T/\Gamma$, $T^* \propto \Gamma$. 
Using the definition of $\Gamma$, we then obtain 
\beq
T^* = \frac{v_F}{a} f_\xi(\g)
\label{4}
\eeq
where $f_\xi(\g)$ is a function of $\g$ and $\xi$.

We also consider the mass renormalization $Z_k (\omega) = 1 + \Sigma_k (\omega)/\omega$, 
which effectively measures the strength of the coupling along the FS. It is given by
\beq
Z_k (\omega) = 1 + \frac{\g T}{a} \sum_{\omega'} \oint dk' \nu_{k'}~
\frac{{\text sign} \omega'}{\omega} D_{k-k'} (\omega-\omega')
\label{3}
\eeq 
One can easily verify that
$Z_k (\pi T) = 1 + (\g/(a \pi)) \oint dk' \nu_{k'}D_{k-k'} (0)$ 
is independent of the Landau damping. For large $\g$, $Z \gg 1$ along the entire FS;
for small $\g$, but large $\xi$, it is still large near the hot spots $k=k_h$,
 where $Z_{k_h} (\pi T) = 1 + \g (\xi/a)$.
 We verified that for $\xi \gg a$, 
fermions relevant for pairing  have $Z_k >1$, i.e.,
pairing in the critical regime
is a strong coupling phenomenon for all $\g$.    

\begin{figure}
\includegraphics*[width=0.9\columnwidth] {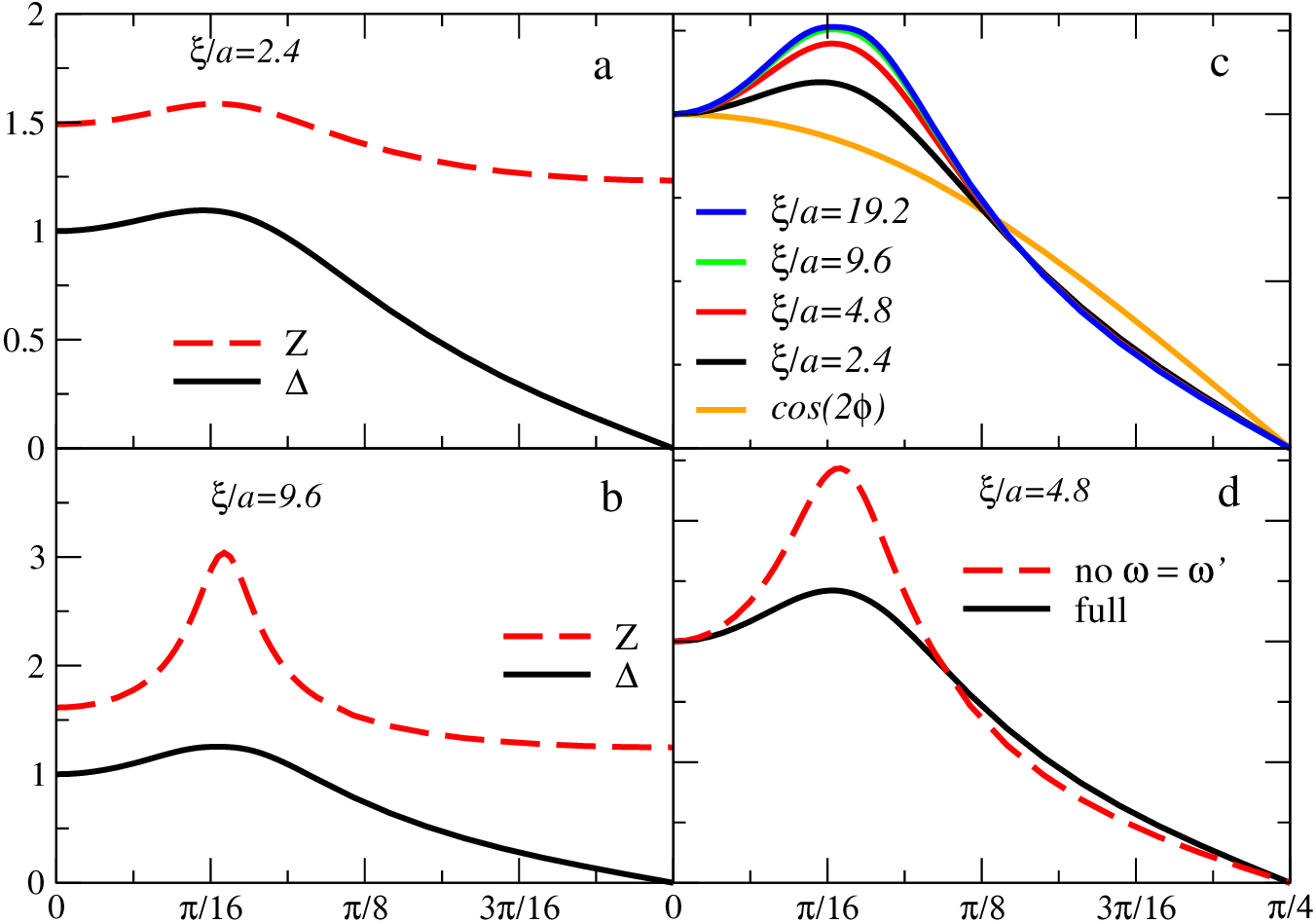}
\caption{(Color online) The mass renormalization
$Z_k (\pi T)$ and the pairing gap $\Delta_k (\pi T)$ for small $\g = 0.2$.
Panels (a) and (b)  --- the results for two different $\xi$;  panel (c) -- the angular dependence of 
the gap compared to the $\cos(k_xa) - \cos(k_ya)$ form; panel (d) -- 
$\Delta_k (\pi T)$ for $\xi =4.8 a$ with and without the pairbreaking contribution  from the static spin 
fluctuations (the term with $\omega = \omega'$ in Eq.~(\protect\ref{1})).
 Without the static contribution, $Z_k (\pi T) \equiv 1$.}
\label{fig2}
\end{figure}

\begin{figure}
\includegraphics*[width=0.9\columnwidth] {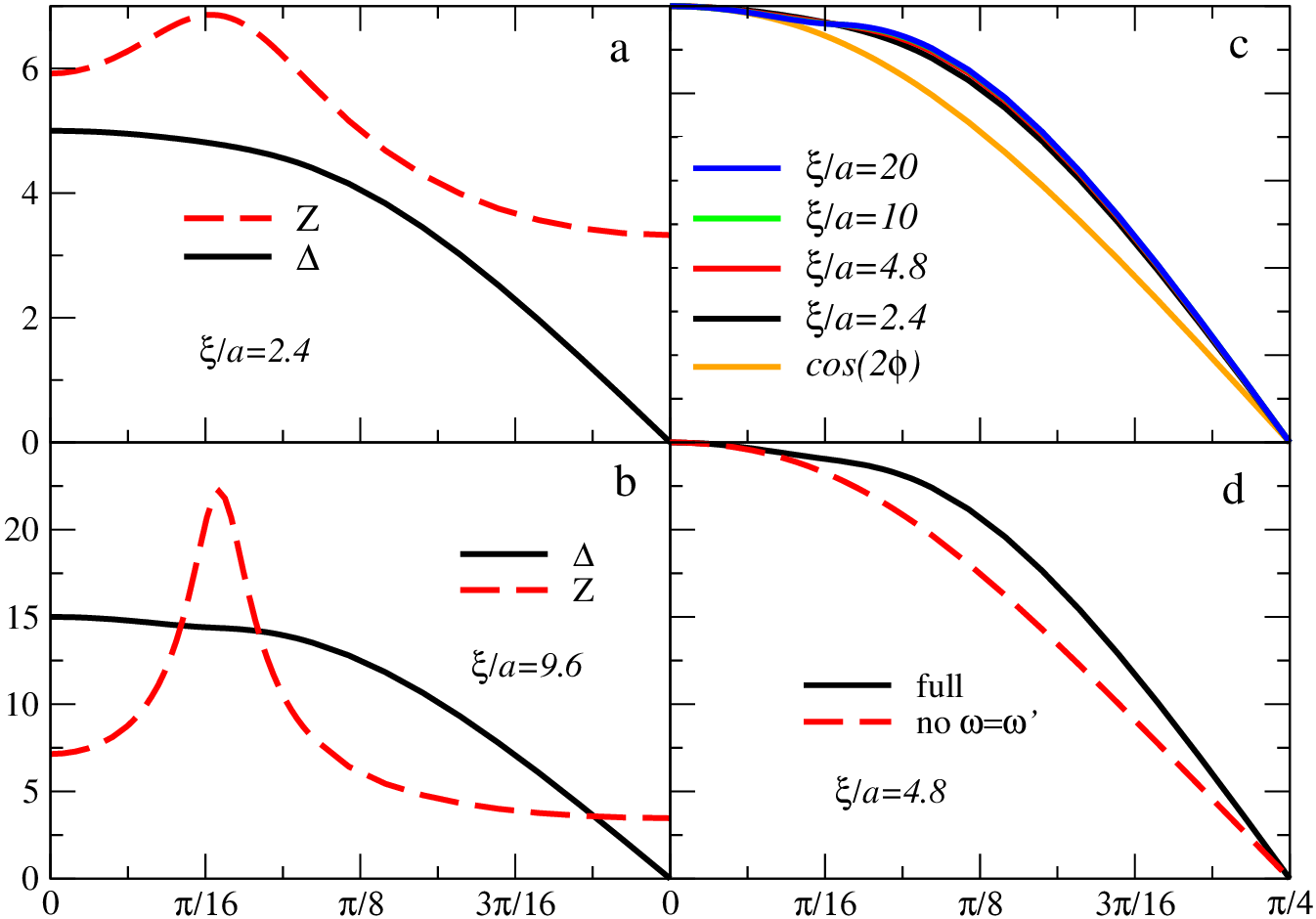}
\caption{(Color online)
Same as Fig.~\protect\ref{fig2}, but for large $\g =2$.} 
\label{fig3}
\end{figure}

{\it The results}~~~The results of the numerical calculations for $T^* (\g)$, 
$\Delta_k (\omega)$ and $Z_k (\omega)$  for the O-Z form of $D_q (0)$ 
are presented in Figs.~\ref{fig1}-\ref{fig3} 
(we used 142 Matsubara frequencies).  $\Delta_k (\omega)$ 
monotonically decreases with increasing frequency (not shown).
The dependences of $\Delta_k (\omega)$ and $Z_k (\omega)$ for $k$ along the FS
 are similar for different $\omega$,
 so we only present the results for the lowest Matsubara frequency $\omega = \pi T$. 

In Fig.~\ref{fig1} we show the dependence of $T^*$ on the dimensionless coupling $\g$. 
We see that $T^*$ initially increases with $\g$, passes through a 
broad maximum at $\g \sim 1$, and then decreases, eventually as $1/\g$. 

In Fig.~\ref{fig2} we show the results for $Z_k (\pi T)$ and $\Delta_k (\pi T)$ for small $\g =0.2$.
We see that $Z_k (\pi T)$ is enhanced near the hot spots and then drops
to near its non-interacting value ($Z=1$), which implies
 that only the hot regions are relevant for pairing.
 $\Delta_k (\pi T)$ also has a maximum at the hot spots, leading to a significant deviation
from the $\cos(k_xa) - \cos(k_ya)$ form.  Moreover,
 the slope of  $\Delta_k (\omega)$ near the node is 
 smaller than this form. 
 
In  Fig.~\ref{fig3} we show the results for large $\g =2$. 
We see that $Z$ is enhanced along the entire FS.
The deviation of the gap anisotropy from the $\cos(k_xa) - \cos(k_ya)$ form is much
 weaker than at small $\g$, and
 the slope of  $\Delta_k (\omega)$ near the node is 
larger than this form. 

\begin{figure}
\includegraphics*[width=0.7\columnwidth] {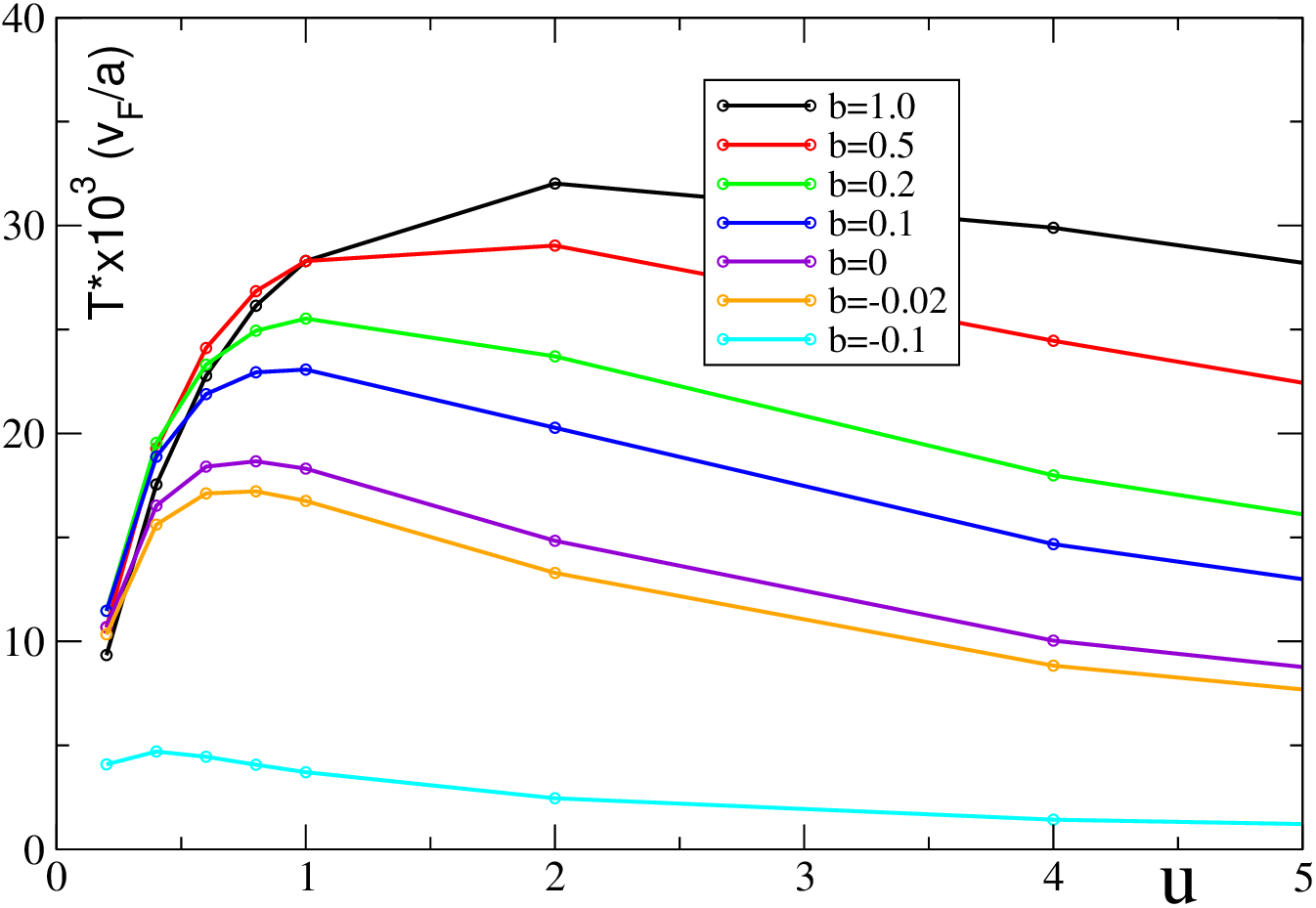}
\caption{(Color online) The onset temperature $T^*(\g)$ for different forms of the static 
susceptibility $D_q (0)$ (see text).
The magnitude of $T^*(\g)$ increases as $D_q(0)$ drops faster with deviation from $Q$.}
\label{fig4}
\end{figure}

In Fig.~\ref{fig4} we show how $T^*(\g)$ changes with the deviation of the static susceptibility 
from the O-Z form. We see that the
 magnitude of $T^*(\g)$ increases for $b>0$, i.e., when 
 $D_q(0)$ drops faster from $Q$ than the O-Z form.

{\it Analytical reasoning}~~~All these results can  be understood analytically. 
Consider first 
why $T^*$ remains finite when $\xi$ diverges, despite the fact that thermal  spin fluctuations 
 are pairbreaking for a $d-$wave gap.
 We recall that Eliashberg theory is a set of two coupled equations for the pairing vertex $\Phi (\omega)$ and 
 self-energy $\Sigma (\omega)$. They reduce to independent equations for $\Delta (\omega)$ and $Z(\omega)$ 
 by exploiting the definitions $\Phi (\omega)  = \Delta (\omega) Z(\omega)$ and 
 $\Sigma (\omega)= \omega (Z(\omega)-1)$. The 
static thermal contribution to Eq.~(\ref{1}) (the term with $\omega = \omega'$) 
 has the form 
\beq
 - \frac{\g T}{a |\omega|} \oint d k' \nu_{k'}~\frac{\Delta_{k'}
(\omega) + \Delta_k (\omega)}{\xi^{-2} + |{\bf k}-{\bf k}'-{\bf Q}|^2}
\label{1_1}
\eeq
where the terms with $\Delta_k$ and $\Delta_{k'}$ are the 
contributions from $\Sigma$  and $\Phi$, respectively.  
Each diverges at $\xi = \infty$, and taken alone, 
would drive $T^*$ to zero. However, the 
 sum of the two remains finite because by symmetry $\Delta_{k+Q} = -\Delta_k$,
  and $T^*$ does not vanish at $\xi = \infty$.  
Note that there is no such cancellation for $Z(\omega)$, which diverges at a hot spot when $\xi =\infty$ 
%AC
(Ref. \onlinecite{new_comm}) 

Consider next  the dependence of $T^*$ on $\g$ for $\xi = \infty$.
  At small $\g$, pairing is confined to hot regions 
($k \approx k_h$), and  the momentum dependence of 
the static $D_{k-k'}$ can be approximated by 
 $|{\bf k}_h-{\bf k}'|^2$. The integral over $k'$ in Eq.~(\ref{1})  can then be 
evaluated analytically, and Eq.~(\ref{1})
 reduces to a one-dimensional integral equation for $\Delta_{k_h} (\omega)$.
Simple power counting then yields $T^* \sim (\Gamma/a^2) \g^2$. 
 A numerical solution gives~\cite{aga}
 $T^* \approx 0.68  (\Gamma/a^2) \g^2 = 0.13 (v_F/a) \g$, i.e.,
 $f_\infty(\g \ll 1) \approx 0.13 \g$.
 
The reduction of $T^* (\g)$ at large $\g$ 
is peculiar to $d-$wave pairing.  The argument is that,
 as $\g$ increases, $T^*$  initially also increases, and  at some $\g$
the dynamic term in $D_{k-k'} (\omega - \omega')$ 
 becomes comparable to a typical $|{\bf k}-{\bf k}' -{\bf Q}|^{2}$ term
along the FS, which determines
 the attractive $d-$wave component of the static interaction. 
A further increase of $T$ would make the effective interaction less momentum dependent 
and hence would reduce the $d-$wave attraction. 
 The balance is reached when $T^*/\Gamma$ is a constant, i.e.,
 when $f_\infty (\g) \propto 1/\g$. Numerically, $f_{\infty} (\g) \sim 0.056/\g$ for the O-Z form of 
 $D_q (0)$. This  can be re-expressed as $T^* \sim 0.47 (v_F/a)^2/U \sim 
0.5 J$, where $J \approx (v_F/a)^2/U$ is the exchange integral
 of the underlying Heisenberg model at half filling~\cite{comm_aaaa}. 
 We interpret this as evidence that 
pairing of incoherent fermions 
  in the  spin-fermion model, and the creation of singlet pairs upon 
 doping a Mott-Heisenberg insulator, describe the same physics from different perspectives. 

At the same time, we argue that, even at strong coupling, pairing is 
 confined to the vicinity of the FS.
 To see this, from the integral over $\epsilon_k$
which leads to Eq.~(\ref{1}), we can estimate a
  typical $|{\bf k}-{\bf k}_F|$ transverse to the FS from
  $v_F |{\bf k}-{\bf k}_F| \sim \omega Z(\omega)$, where $\omega \sim \pi T^*$ is a typical frequency for the pairing 
  problem.  Using the quantum critical form~\cite{adv}
$Z(\omega) \approx 2 \g \sqrt{\Gamma/\omega}$,
 we find that for large $\g$,
 $|{\bf k}-{\bf k}_F| \sim \omega_{typ} Z(\omega_{typ})/v_F \sim (3/8) \sqrt{\pi T^*/\Gamma} \sim 0.1 (\pi/a)$.
 This $|{\bf k}-{\bf k}_F|$ is numerically much smaller than 
$k_F \sim 0.8 (\pi/a)$, i.e., pairing 
  involves fermions  from a narrow shell around the FS. For smaller $\g$,  $|{\bf k}-{\bf k}_F|/k_F$ is even smaller.

Consider next the variation of the gap along the FS. 
 For small $\g$, the anisotropy of the gap is a
 consequence of the fact that the pairing problem is confined to a region near the
hot spots with width $\delta k \sim k_F \g$.
We found that the ratio of the slope of the gap near the node to the gap value at the hot spots 
scales as $\g^2$ for small $\g$.  
At strong coupling,  
 the non-confinement of $\Delta_k$ to hot spots is due to 
 a cancellation between the anisotropies of $\Phi$
and $Z$ when dividing to form $\Delta$. 
 Further, expanding Eq.~(\ref{1_1}) 
near ${\bf k}'={\bf k} +{\bf Q}$,   we find that the residual, non-singular pairbreaking contribution from thermal spin fluctuations
 contains $\partial^2\Delta/\partial^2 k'$, 
i.e., it tends to make  the gap more smooth.
  The extent to which this affects the shape of the gap
 can be verified numerically. In the last panels in Figs.~\ref{fig2} and \ref{fig3},
 we show the gap variation along the FS with and without the pairbreaking contribution 
 to the gap equation. 
We see that $\Delta_k$ near the node becomes larger than the $\cos(k_xa)-\cos(k_ya)$ form when the pairbreaking contribution  is included.
   
Finally, we consider the variation of the overall scale for $T^*$ with $b$ 
(Fig.~\ref{fig4}). To understand this, we recall that for the O-Z form of $D_q (0)$, a strong reduction of 
$T^*$ compared to the asymptotic, small $\g$ form $T^* \approx 0.13 (v_F/a) \g$ was due to the interplay 
between the $|\omega -\omega'|/\Gamma$ term  and the maximum $|{\bf k}-{\bf k}'|$  along the FS.
 Once $D_q (0)$ becomes steeper with deviation from $Q$ ($b >0$),  the typical $k$ and $k'$ 
 get closer to the hot spots, the restriction on $|{\bf k}-{\bf k}'|$ becomes less relevant, and  $T^*$ increases.
  For $b<0$, the situation is the opposite, geometrical restrictions become more relevant, and $T^*$ rapidly 
  decreases.

{\it Comparison with earlier studies}~~~Several groups did extensive studies of  $T^*$ within the Eliashberg theory  
for the spin-fermion 
 model~\cite{andy,mp,norman}. They treated $\Gamma$ as an
 independent parameter, i.e., they did not express it in terms of $\g$.
 Our results agree with these studies if we use the same $v_F$, $\g$ and $\Gamma$ as they did.  
 Monthoux and Pines~\cite{mp} studied 
 $T^* (\g)$ at small $\g$ and noticed that its slope decreases as the coupling increases, consistent with 
 Fig.~\ref{fig1}.
  Schuttler and Norman~\cite{norman} 
 found a saturation of $T^*$ when measured in units of $\Gamma$, but
 for a model in which $D(q,\Omega)$ had the factorized form $D_1(q) D_2 (\Omega)$, and a strong $T$ dependence of $\Gamma$ is introduced phenomenologically. Within the spin-fermion 
 model, though, the $T$ dependence of $\Gamma$ is weak. 
Our results also agree with  FLEX calculations
for the Hubbard model~\cite{ms,eremin,carbotte}. In these calculations, 
 $\Gamma$ is obtained self-consistently, as in our theory. For
 $U=4t$ and $v_F/a \approx 2t$ ($\g \approx 0.25$), our 
$T^* \approx 0.02 t$ is in good agreement with Refs.~\onlinecite{ms,eremin,carbotte}. This value is also in good agreement with two-particle self-consistent 
 calculations~\cite{kyung}, dynamical cluster approximation~\cite{maier},  
and cluster DMFT~\cite{haule}, which  also yields 
%AC
 that the $d-$wave order parameter scales as $J$ 
%$T^* \propto J$ 
at large $\g$ (Ref.~\onlinecite{koncharla}).
% as in Fig.~1. 

{\it Comparison with the cuprates}~~~~To get $T^*$ in Kelvin, we use $v_F/a \sim 1 eV$, noting that $v_F$ is the `bare' velocity as obtained in band theory.
 For the O-Z form of the static 
$D_q (0)$, we obtain  $T^* \sim 0.02 v_F/a \sim 200-250K$ 
 for $\g \sim O(1)$.  For $b =1$, this temperature increases to over $350K$ (Fig.~\ref{fig4}), and becomes 
 even larger for larger $b$.  This shows that a spin-mediated pairing interaction is strong 
 enough to account for experimental values of $T^*$.

We also verified that the `hot spot' description, which yields a very anisotropic $d-$wave gap, is only valid for
small $\g$,  
which would correspond to strongly overdoped cuprates.
But in this case, $\xi$ is also
 small, and therefore the enhancement of $\Delta_k$ near the hot spots is weakened.
For optimal and underdoped cuprates, we find a gap close to the $\cos(k_xa) - \cos(k_ya)$ form. This is consistent 
with photoemission~\cite{kanigel} below $T_c$, and with the scenario that the Fermi arcs above $T_c$ appear 
because of thermal broadening of the spectral function~\cite{mikearcs}.  

{\it Conclusions}~~~~In this paper, we analyzed how  $d_{x^2-y^2}$ pairing
mediated by nearly critical
spin fluctuations varies with the coupling strength.
We argued that the onset temperature for pairing $T^* \approx (v_F/a) f_\xi (\g)$
 smoothly evolves between weak and strong couplings, passing through a shallow maximum 
 at $\g \sim 1$.
 At large $\g$,  $T^* \sim v_F/\g \sim J$.
For all $\g$, pairing is confined to the vicinity of the FS.
 We also argued that $T^*$ only weakly depends on the distance to the antiferromagnetic instability, as singular 
pairbreaking 
contributions from static spin fluctuations cancel out in the gap equation. The residual
 pairbreaking terms only modestly reduce $T^*$, and at the same time smooth the angular dependence 
 of the gap, and at strong coupling
make $\Delta_k$ near the node even larger  than the $\cos(k_xa) - \cos(k_ya)$ form. 
As Mott physics is certainly present near half filling, where $T^*$ reaches 
 its largest values, our $T^* (\g)$ should only be taken as an estimate. 
 Still, the fact that $T^*$ is in the experimental range is in support
of a one-gap scenario in which  
 the  instability at $T^*$ occurs in the particle-particle channel, due to interactions with  spin fluctuations.

%AC
We acknowledge helpful discussions with M. Eschrig, I. Eremin,
M. Jarrell, G. Kotliar, A. J. Millis,  A.-M. Tremblay, D. Scalapino, 
and I. Vekhter,  support
from the Welch foundation (ArA),  NSF-DMR 0604406 (AVC) and
the US DOE, Office of Science, Contract No.~DE-AC02-06CH11357 (MRN).

 \end{document}